\documentclass[10pt]{article}
\usepackage[utf8]{inputenc}

\usepackage[letterpaper,margin=1in]{geometry}
\usepackage[colorlinks,allcolors=blue]{hyperref}
\usepackage[parfill]{parskip}
\usepackage{amsthm,amsmath}
\usepackage[round]{natbib}
\usepackage{setspace}
\usepackage{pgf,tikz}
\usetikzlibrary{positioning}
\usepackage{authblk}

\newtheorem{proposition}{Proposition}

\theoremstyle{definition}

\newcommand{\indep}{\protect\mathpalette{\protect\independenT}{\perp}}
\def\independenT#1#2{\mathrel{\rlap{$#1#2$}\mkern2mu{#1#2}}}

\title{Comment: Reflections on the Deconfounder}
\author{Alexander D'Amour \\ Google Research, Cambridge, MA, USA \\ alexdamour@google.com}
\date{}

\begin{document}

\maketitle

I would like to congratulate the authors on their illuminating article, and thank the editors for the opportunity to discuss the paper. 
The deconfounder method that this article presents is appealing: a number of important scientific investigations and high-stakes decisions fit into its template.
Indeed, as the authors note, instances of the deconfounder have already been deployed without explicit causal language in a number of applied settings.
By bringing to light the implicit causal argument that underlies this approach, the authors have sparked an important conversation with potentially far-reaching consequences.
It is thus important to carefully outline when we expect the deconfounder method to succeed in characterizing causal relationships and when we expect it to fail.

I have personally been in conversation with the authors over the past two years about this work, and this discussion has yielded some interesting insights, some of which have been published \citep{d2019multi}, and some of which now appear in the current version of the article and in follow-up work \citep{wang2019multiple}.
The aim of this note is to draw out some conclusions from this conversation about the role that the deconfounder can play in practical causal inference.
In particular, I will make three points here.
First, in my role as the critic in this conversation, I will summarize some arguments about the lack of causal identification in the bulk of settings where the ``informal'' message of the paper suggests that deconfounder could be used.
This is a point that is discussed at length in \citet{d2019multi}, which motivated the results concerning causal identification in Theorems 6--8.
Second, I will argue that adding parametric assumptions to the working model in order to obtain identification of causal parameters (a strategy followed in Theorem 6 and in the experimental examples) is a risky strategy, and should only be done when extremely strong prior information is available.
Finally, I will consider the implications of the nonparametric identification results provided for a narrow, but non-trivial, set of causal estimands in Theorems 7 and 8.
I will highlight that these results may be even more interesting from the perspective of \emph{detecting} causal identification from observed data, under relatively weak assumptions about confounders.

Throughout this note, I will draw connections to sensitivity analysis methods that probe the implications of unobserved confounding.
This is a natural lens through which to study the deconfounder because many sensitivity analysis methods posit a similar latent variable model to the one that the deconfounder deploys as a working model \citep[see, e.g., ][]{rosenbaum1983assessing}.
Well-designed sensitivity analyses can reveal how specific assumptions restrict the range of causal conclusions that are compatible with the observed data, and are thus useful for understanding what is lost when assumptions like ``no unobserved confounders'' are relaxed to ``no unobserved single-cause confounders.''
Thus, I believe, as the authors suggest, that sensitivity analysis should be a core part of any workflow that deploys the deconfounder, and discuss at various places how sensitivity analysis could be used effectively in this setting.

\paragraph{Preliminaries} Following the paper, I will denote causes as $A := (A^{(1)}, \ldots, A^{(m)})$ taking specific values $a = (a^{(1)}, \ldots, a^{(m)})$, potential outcomes as $Y(a)$, and latent confounders as $Z$.
To avoid measure-theoretic considerations when writing conditioning statements, I will consider the treatments $A^{(k)}$ to be discrete.
I will write observed outcomes as $Y^{obs}$, where, under the stable unit treatment value assumption (SUTVA), $Y^{obs} = Y(A)$.
Finally, I will denote by $Z$ any latent confounders.

Throughout, I will consider models of the joint distribution $P(A, Y^{obs}, Z)$, which I will refer to as latent variable models.
I will assume that unconfoundedness is satisfied conditional on $Z$:
    \begin{equation*}
    Y(a) \indep A \mid Z \quad Z\text{-a.e.}, \forall a.
    \end{equation*}
Thus, if the latent variable model is fully specified, the potential outcome distributions $P(Y(a))$ are also specified by the following adjustment formula, which ``adjusts'' for the confounder $Z$ 
    \begin{equation}\label{eq:adjustment}
    P(Y(a)) = E[P(Y^{obs} \mid Z, A = a)] \quad \forall a.
    \end{equation}
I will refer to the integrand in \eqref{eq:adjustment} $P(Y^{obs} \mid Z, A = a)$ as the outcome model.
If the confounder $Z$ is observed, and the overlap condition is satisfied, then $P(Y(a))$ is identified from observed data. 
The question at hand is whether $P(Y(a))$ can be identified when $Z$ is unobserved.

\section{Fundamental Limitations of the Deconfounder Approach}

I will begin by summarizing the argument in \citet{d2019multi} critiquing the ``informal'' message about the deconfounder approach (stated most explicitly in the informal statement of Theorem 6 and Section 3.4).
Specifically, this message asserts that, under the ``no unobserved single-cause confounders'' assumption, any well-fitting latent variable model $P(Y^{obs}, A, Z)$ will yield the correct potential outcome distribution in $P(Y(a))$ via the adjustment formula \eqref{eq:adjustment}.
This informal story is motivated by strong intuition.
Lemmas 1--3 establish that multi-cause confounding leaves an observable ``imprint'' of dependence between the causes $A$.
Thus, it seems natural that we might be able to gain some information, and even adjust for, an unobserved multi-cause confounder $Z$ by modeling the dependence between the causes $A$.

Unfortunately, this intuition can only be carried so far: while a factor model for the causes $A$ can recover information about multi-cause confounders from observed data, the potential outcome distributions $P(Y(a))$ are not non-parametrically identified, except in cases where all confounding is observed.
Thus, without additional unverifiable assumptions, no method can recover the distributions $P(Y(a))$ when there is unobserved confounding.
In this section, I briefly demonstrate why this is the case.
For a more in-depth argument about lack of identification in this setting with concrete examples, see \citet{d2019multi}.

As I show formally below, the key difficulty is that the causes $A$ cannot be used simultaneously as measurements of the unobserved confounder $Z$, and as treatments whose effects are being estimated.
If the event $A = a$ provides only a noisy measurement of $Z$, there is ambiguity in how the outcome model $P(Y^{obs} \mid Z, A = a)$ should align the variability in the residual distributions $P(Y^{obs} \mid A= a)$ and $P(Z \mid A = a)$; there are many specifications of the residual dependence between $Y^{obs}$ and $Z$ that are compatible with the observed data.
This is a classic problem that arises when confounders are measured with error \citep[see, e.g.][]{ogburn2012bias}.
On the other hand, if the event $A = a$ provides a perfect measurement of $Z$, such that there is some function $\hat z(A)$ such that $\hat z(a) = Z$, then the overlap condition fails.
In this case, $P(Y^{obs} \mid Z, A = a)$ is only identified when $Z = \hat z(a)$ because the event $Z \neq \hat z(a)$ has zero probability in the observed data.

Let us now make this argument formal.
To do this, we will account for how the two deconfounder assumptions of (a) good model fit, and (b) ``no unobserved single-cause confounders'' constrain the factor model and its implications about the potential outcomes $P(Y(a))$.
This accounting is convenient if we rewrite the joint distribution using copula densities $c(V, W) = \frac{P(V, W)}{P(V)P(W)}$, which characterize the dependence between random variables independently of their marginal distributions.
\begin{align}\label{eq:sensitivity decomp}
    P(Y^{obs}, A, Z) &= \underbrace{P(A, Y^{obs})}_{\text{Observed}} \cdot
    \underbrace{P(Z) c(Z, A)}_{\text{Factor Model}} \cdot
    \underbrace{c(Y^{obs}, Z \mid A)}_{\text{Outcome Copula}}.
\end{align}
Each factor in this composition corresponds to a different assumption.
The requirement for good model fit constrains only the first term, which specifies the distribution of observable quantities, while the ``no unobserved single-cause confounders'' assumption constrains the second term by constraining the causes to be conditionally independent given $Z$ (Lemma 2).
\footnote{The ``no unobserved single-cause confounders'' assumption does not uniquely identify the factor model by itself. Some structure also needs to be put on the latent variable, and even then, the factor model may not be identified. See \citet{d2019multi} for an example where the factor model $P(A, Z)$ is itself not identified.} 
This leaves the outcome-confounder copula density $c(Y^{obs}, Z \mid A) = \frac{P(Y^{obs}, Z \mid A)}{P(Y \mid A)P(Z \mid A)}$ unconstrained.
This copula specifies the residual dependence between $Y^{obs}$ and $Z$ after conditioning on the causes $A$, and plays a key role in specifying the outcome model $P(Y^{obs} \mid A, Z)$.

To complete the argument, note that the potential outcome distributions $P(Y(a))$ implied by the latent variable model are sensitive to the specification of this copula.
Specifically, the estimand in \eqref{eq:adjustment} can be written as
$$
P(Y(a)) = \int_{Z} P(Y^{obs} \mid A = a) c(Y^{obs}, Z \mid A = a) dP(Z).
$$
Plugging in different specifications of the copula here yields different conclusions about $P(Y(a))$.
Whenever $P(Y(a)) \neq P(Y \mid A = a)$, there are multiple specifications of the copula that yield different conclusions about the potential outcomes.
\footnote{To see this, note that the independence copula $c(Y^{obs}, Z \mid A = a) = 1$ implies that $P(Y(a)) = P(Y \mid A = a)$. Thus, because $P(Y(a)) \neq P(Y \mid A = a)$, this copula and the true copula yield different conclusions about $P(Y(a))$.}
Thus, $P(Y(a))$ is not identified unless there is no confounding and $P(Y(a)) = P(Y \mid A = a)$.

We can now revisit the tension between the roles of causes $A$ as measurements of $Z$, and as treatments. 
In cases where $Z$ can only be inferred inexactly (i.e., $P(Z \mid A = a)$ is non-degenerate), the marginals $P(Y^{obs} \mid A = a)$ and $P(Z \mid A = a)$ put some constraints on the outcome model $P(Y^{obs} \mid Z, A = a)$, but the ambiguity in the copula implies that this model is not identified for any value of $Z$.
In cases where $Z$ can be reconstructed deterministically from the causes by some function $\hat z(a)$, (i.e., $P(Z \mid A = a)$ is degenerate), the outcome model $P(Y^{obs} \mid Z, A=a)$ is identified when $Z = \hat z(a)$, but the copula is undefined whenever $Z \neq \hat z(a)$ because this event has zero probability.

The upshot of this argument is that neither the deconfounder nor any other estimation method can adjust for unobserved confounding when estimating $P(Y(a))$ under the ``no unobserved single-cause confounders'' assumption alone.
This conclusion holds no matter how much information we can glean about an unobserved confounder $Z$ from the causes $A$.
Although the single-cause confounding assumption does put some non-trivial structure on the latent variable model, it is not enough for causal estimation.

This lack of identification leaves practitioners looking to apply the deconfounder with two options: either make additional assumptions about the latent variable model $P(Y^{obs}, A, Z)$ so that $P(Y(a))$ is identified, or seek out causal comparisons where all of the confounding is effectively observed.
In the Theory section of the paper, the authors consider both of these paths.
I will discuss each of these options in turn.

\section{Parametric Identification, If You Must}

I now turn to the subject of parametric identification of causal parameters, and offer some cautions about employing this strategy.
Parametric identification is a natural strategy to employ when the causal parameters of interest are not non-parametrically identified. 
One obtains parametric identification by adding parametric assumptions to the working model that constrain the implied potential outcome distributions $P(Y(a))$ to be unique.
The authors employ this parametric identification strategy in the experimental demonstrations of the deconfounder, as well as the formal result in Theorem 6.
In Theorem 6, the copula $c(Y^{obs}, Z \mid A)$ is restricted by assuming that there is no interaction between the causes $A$ and the latent variable $Z$ in the outcome model (i.e., that they combine linearly), and assuming that the confounder is piecewise constant in $A$.
In the paper's experiments, the authors assume a parametric factor model (e.g., a quadratic factor model for the genome-wide association study simulation), and a true linear outcome model.
In the cases of Theorem 6 and the GWAS simulation study, the authors prove that these parametric assumptions are sufficient for identification.

Parametric identification can be a risky strategy to employ in practice.
Specifically, the fact that the parametric assumptions are necessary to identify causal parameters implies that some aspects of these assumptions are not testable in the observed data.
The decomposition in \eqref{eq:sensitivity decomp} makes this clear: given that the observed data are insufficient to identify the causal parameters, the parametric assumptions must restrict some of the unidentified portions of the latent variable model.
Thus, to have confidence in this approach, one needs to have confidence in the parametric model used to identify causal effects as a \emph{true model of the world}, not merely as an acceptable description of the observed data.
This is because the identifying parametric assumptions specify not only a descriptive model of the observed data, but also a structural model for unobserved counterfactual outcomes.
Relying on parametric identification may be feasible in cases where one has strong prior knowledge---e.g., about the quantity represented by the unmeasured confounder, or the specific distributions of measurement errors---but such knowledge is often unavailable.

In addition, uncertainty estimates that are based directly on the parametric specification, e.g., Bayesian credible sets, do not capture the full extent of uncertainty about causal effects according to the data.
Specifically, these uncertainty estimates only quantify uncertainty \emph{within} the specified model, and do not include the fundamental uncertainty associated with the lack of non-parametric identification of the potential outcome distributions $P(Y(a))$.
As a result, unless the prior information used to specify the parametric assumptions is very strong, these uncertainty estimates will understate the degree of uncertainty about a causal parameter estimate.
This is a standard critique of parametric uncertainty quantification, but carries extra weight in the context where conclusions depend on untestable aspects of the parametric model.
For example, for the parametrically identified latent variable model in the GWAS example, as the sample size grows, the posterior for the causal parameter will concentrate around a single value, even though there exists a range of outcome models that correspond to different copulas $c(Y^{obs}, Z \mid A= a)$ that are equivalently compatible with the observed data, but would concentrate on different causal parameters.
In fact, even small, seemingly benign parametric choices can mask alternative causal explanations.
Lessons from latent variable models in the missing data and causal inference literatures can be instructive here.
For example, analyses of the widely-used Heckman selection model \citep{heckman1979} have noted that the tail thickness of priors on latent variables can induce starkly different conclusions that are hidden by using the Gaussian default \citep{littlebook,ding2014bayesian}.
See also discussions in \citet{robins2000sensitivity} and \citet{linero2017bayesian} for other examples.

Here, sensitivity analysis can be a useful tool to account for the fundamental uncertainty due to non-identification of the causal estimand. 
When performed with parametric models, sensitivity analyses perturb the parametric assumptions made with the estimating model in order to understand what other causal conclusions could be obtained under different parametric specifications. 
Performing sensitivity analyses on deconfounder estimates is straightforward: a number of sensitivity analysis approaches employ a working model with the same latent variable structure \citep[e.g., ][]{rosenbaum1983assessing,imbens2003sensitivity,dorie2016flexible,cinelli2018making}.
However, sensitivity analyses can also fall victim to spurious parametric identification if the perturbations are not appropriately parameterized \citep{gustafson2018sensitivity}.
To avoid this issue, it can be useful to employ sensitivity analysis strategies that cleanly separate the portions of the model that are identified by the observed data from those that are identified by parametric assumptions \citep{franks2019flexible,robins2000sensitivity,linero2017bayesian}.
In the context of the deconfounder, the decomposition in \eqref{eq:sensitivity decomp} is a promising place to start, and is the subject of current work.

\section{Toward a More Selective Deconfounder Workflow}

A more cautious alternative to pursuing parametric identification is to seek out causal questions that have definitive answers under the ``no unobserved single-cause confounders'' assumption.
The authors take this path in Theorems 7 and 8, in a setting where the latent confounder $Z$ can be deterministically reconstructed as a function of the causes $\hat z(A)$.
Here, however, the factor model seems less interesting as a tool for calculating causal effects, and more interesting as a tool for establishing empirically when no unobserved confounding is present. 
In my opinion, this seems to be a more interesting thread to follow.

To review, in Theorem 7 the authors consider partitioning the causes into a set of focal causes $A_{1:k}$ whose effects will be estimated, and a set of auxiliary causes $A_{k+1:m}$ that will serve as measurements of the latent confounder.
The theorem then states that if the latent confounder $Z$ can be written as a function of the auxiliary causes $Z = \hat z(A_{k+1:m})$ alone,
\footnote{This is not how the theorem is stated, but this function restriction is implied by the subsequent overlap condition.}
then the distributions of potential outcomes defined with respect to the subset of focal causes $P(Y(a_{1:k}))$ are identifiable subject to an overlap condition.
Meanwhile, Theorem 8 states that certain counterfactual potential outcome distributions of the form $P(Y(a) \mid A = a')$ are identifiable as long as the causes $a$ and $a'$ map to the same value of the latent confounder, i.e., $\hat z(a) = \hat z(a')$.

In these results, the authors focus on the role of the factor model in the identification of causal estimands under the ``no unobserved single-cause confounders'' assumption.
However, the factor model is not essential for this point.
Note that Theorems 7 and 8 both imply that the causal parameters can be identified in terms of the causes $A$ alone, because it is assumed that the confounder $Z$ can be written as a function of $A$.
Written with slightly more generality, the identification result in Theorem 7 implies:
\begin{align}\label{eq:thm7}
P(Y(a_{1:k})) = E[P(Y^{obs} \mid A_{1:k} = a_{1:k}, A_{k+1:m})],
\end{align}
while the identification result in Theorem 8 implies: 
\begin{align}\label{eq:thm8}
P(Y(a') \mid A = a) = P(Y^{obs} \mid A = a') \quad \forall (a, a') \text{ s.t. } \hat z(a) = \hat z(a').
\end{align}

To me, the more interesting point is that the factor model can be used in some cases to determine empirically whether some of the assumptions of the theorems are met.
For example, the setting of Theorem 7 can be framed as a problem where the unobserved confounder $Z$ is measured with proxies $A_{k+1:m}$.
It is well-understood that in the limit where $Z$ is perfectly recovered by the proxies, the potential outcome distribution $P(Y(a_{1:K}))$ is identified \citep{ogburn2012bias};
however, in single-cause problems, one cannot determine whether this condition has been met.
Similarly, Theorem 8 can be framed as a setting where one is imputing a set of counterfactual outcomes within a subpopulation where there is no confounding because, within this subpopulation, the confounder is fixed.
Here, too, in single-cause problems, one cannot definitively identify such subpopulations from observed data. 
Interestingly, the theory of multi-cause confounding presented in the paper suggests that these assumptions can be empirically validated under some restrictions on the causal DAG relating $A$ to $Y^{obs}$ and the ``no unobserved single-cause confounders'' assumption.
For example, this theory supports the following proposition. 

\begin{figure}
\centering
\begin{tikzpicture}[var/.style={draw,circle,inner sep=0pt,minimum size=0.8cm}]
    \node (latent) [var, fill=black!10] {$Z$};
    \node (outcome) [var, below right=0.5cm and 0.5cm of latent] {$Y^{obs}$};
    \node (causem) [var, below left=0.5cm and 0.5cm of latent] {$A^{(m)}$};
    \node (dots) [left of=causem] {$\cdots$};
    \node (cause1) [var, left of=dots] {$A^{(1)}$};
    \node (covariates) [var, right=0.75cm of latent] {$X$};
    
    \path[->]
        (latent) edge (outcome)
                 edge (cause1)
                 edge (causem)
                 edge (covariates)
        (cause1) edge [bend right] (outcome)
        (causem) edge (outcome)
        (covariates) edge (outcome);
\end{tikzpicture}
\caption{DAG assumed in Proposition~\ref{prop:testable}, representing the relationship between causes $A$, latent confounder  $Z$, covariates $X$, and observed outcome $Y^{obs}$.\label{fig:dag}}
\end{figure}
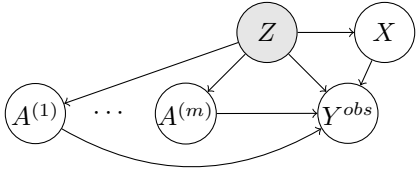

\begin{proposition}\label{prop:testable}
Suppose there are no single-cause confounders, and the structural relationships between causes $A$, latent confounder   $Z$, and observed outcomes $Y^{obs}$ can be represented in the DAG in Figure~\ref{fig:dag}.
Suppose that in addition to causes $A$, we also have auxiliary covariates $X$, which are conditionally independent of the causes $A$ conditional on the multi-cause confounder $Z$.
Then for any function $\hat z(A, X)$ such that the causes $A$ are mutually independent conditional on $\hat z(A, X)$, the conditional independence $A \indep Y(a) \mid \hat z(A, X)$ also holds for each $a$.
\end{proposition}

Theorems 7 and 8 can be written as consequences of this proposition.
This proposition is potentially useful because it shows that absence of certain confounding structures has observable implications.
This insight is closely related to the literature on negative controls \citep[see, e.g.,][]{lipsitch2010negative}.

This result suggests that one can use a similar workflow to the deconfounder to determine, at least in principle, whether identification statements like \eqref{eq:thm7} or \eqref{eq:thm8} are valid in a given setting.
Specifically, one can obtain a function $\hat z(A, X)$ (perhaps by fitting a factor model), then test whether the causes $A$ appear to be mutually independent conditional on $\hat z(A, X)$.
If one is satisfied that this is true, \eqref{eq:thm7} or \eqref{eq:thm8} can be applied.
Importantly, this procedure is truly agnostic to the parametric specification of the model used to obtain $\hat z(A, X)$: all of the conditions are only functions of observables.

While the workflow in this procedure is similar to the deconfounder, it has a different use case.
Instead of enabling causal inference in a wide range of cases, this procedure would be used to determine whether one can proceed with unconfounded inference at all, and can potentially give ``no'' as an answer.
Still, this sort of procedure can prove useful in complex data contexts, where it can be valuable to surface causal questions that can be adequately answered with the available data.
In a specific example of this approach, \citet{sharma2018split} propose a similar testing procedure to uncover unconfounded comparisons, and use it to evaluate the causal effect of a recommender system on purchasing rates for certain products.

In outlining this procedure, I have belabored the point that it is a workflow ``in principle'' because it could prove tricky to implement.
The observable implication that needs to be tested is a complex conditional independence statement, and these are notoriously difficult to test in practice \citep{shah2018hardness}.
In particular, one would receive the ``green light'' to estimate a causal parameter by failing to reject the null of conditional independence, which can only be reliably depended upon if the test has acceptably high power, but designing such tests is difficult, and in some settings, impossible.

Here, it can again be helpful to turn back to sensitivity analysis.
Instead of attempting to rule out all possible forms of dependence between the causes $A$ conditional on $\hat z(A, X)$, a sensitivity analysis approach could explore a number of candidate models for the residual dependence between the causes $A$ and relate these models to the confounding induced by the unobserved confounder $Z$.
For example, one could examine the range of causal effects that would be compatible with the assumption that, conditional on $\hat z(A, X)$, the the causes $A$ are no more predictive of a potential outcome $Y(a)$ than any leave-one-out set of the causes $A_{-k}$ is able to predict a held-out cause $A^{(k)}$. 
This sort of calibration argument is common in more standard sensitivity analyses \citep{imbens2003sensitivity,dorie2016flexible,franks2019flexible,cinelli2018making}.
In cases where dependence between the causes can be ruled out conclusively, this approach would yield a sensitivity region that collapses to a point; however, in the more likely case where many dependences cannot be ruled out, this approach would represent this uncertainty with a wider sensitivity region.
It should be noted that constructing a plausible sensitivity analysis of this type would require deep domain knowledge to justify the analogy between different dependences between variables.
Negative control methods and related identification strategies \citet{lipsitch2010negative} and \citet{miao2018identifying} could be framed as particularly successful executions of this type of argument.

\section{Conclusion}
In writing this paper, the authors have drawn attention to a problem that is simultaneously scientifically important, methodologically interesting, and conceptually subtle.
Although I have taken on the role of critic in our conversations, I believe their contribution here is important.
I remain skeptical about the deconfounder as a method for causal point estimation, but believe that the authors' characterization of multi-cause confounding could yield fruitful developments in sensitivity analysis, and in potentially obtaining identification results in more complex settings.
This work has certainly inspired me to pay more attention to this problem, and to consider how new methods and tools can be developed to help practitioners draw principled causal conclusions in this setting.

\singlespacing
\bibliographystyle{plainnat}
\bibliography{blessings_comment}

\begin{thebibliography}{18}
\providecommand{\natexlab}[1]{#1}
\providecommand{\url}[1]{\texttt{#1}}
\expandafter\ifx\csname urlstyle\endcsname\relax
  \providecommand{\doi}[1]{doi: #1}\else
  \providecommand{\doi}{doi: \begingroup \urlstyle{rm}\Url}\fi

\bibitem[Cinelli and Hazlett(2018)]{cinelli2018making}
Carlos Cinelli and Chad Hazlett.
\newblock Making sense of sensitivity: Extending omitted variable bias.
\newblock Technical report, Working Paper, 2018.

\bibitem[Ding(2014)]{ding2014bayesian}
Peng Ding.
\newblock Bayesian robust inference of sample selection using selection-t
  models.
\newblock \emph{Journal of Multivariate Analysis}, 124:\penalty0 451--464,
  2014.

\bibitem[Dorie et~al.(2016)Dorie, Harada, Carnegie, and
  Hill]{dorie2016flexible}
Vincent Dorie, Masataka Harada, Nicole~Bohme Carnegie, and Jennifer Hill.
\newblock A flexible, interpretable framework for assessing sensitivity to
  unmeasured confounding.
\newblock \emph{Statistics in medicine}, 35\penalty0 (20):\penalty0 3453--3470,
  2016.

\bibitem[D’Amour(2019)]{d2019multi}
Alexander D’Amour.
\newblock On multi-cause causal inference with unobserved confounding:
  Counterexamples, impossibility, and alternatives.
\newblock In \emph{The 22nd International Conference on Artificial Intelligence
  and Statistics}, pages 3478--3486, 2019.

\bibitem[Franks et~al.(2019)Franks, D’Amour, and Feller]{franks2019flexible}
Alex Franks, Alex D’Amour, and Avi Feller.
\newblock Flexible sensitivity analysis for observational studies without
  observable implications.
\newblock \emph{Journal of the American Statistical Association}, \penalty0
  (just-accepted):\penalty0 1--38, 2019.

\bibitem[Gustafson et~al.(2018)Gustafson, McCandless,
  et~al.]{gustafson2018sensitivity}
Paul Gustafson, Lawrence~C McCandless, et~al.
\newblock When is a sensitivity parameter exactly that?
\newblock \emph{Statistical Science}, 33\penalty0 (1):\penalty0 86--95, 2018.

\bibitem[Heckman(1979)]{heckman1979}
James~J Heckman.
\newblock Sample selection bias as a specification error.
\newblock \emph{Econometrica}, 47\penalty0 (1):\penalty0 153--161, 1979.

\bibitem[Imbens(2003)]{imbens2003sensitivity}
Guildo~W Imbens.
\newblock Sensitivity to exogeneity assumptions in program evaluation.
\newblock \emph{American Economic Review}, 93\penalty0 (2):\penalty0 126--132,
  2003.

\bibitem[Linero and Daniels(2017)]{linero2017bayesian}
Antonio~R Linero and Michael~J Daniels.
\newblock Bayesian approaches for missing not at random outcome data: The role
  of identifying restrictions.
\newblock 2017.

\bibitem[Lipsitch et~al.(2010)Lipsitch, Tchetgen, and
  Cohen]{lipsitch2010negative}
Marc Lipsitch, Eric~Tchetgen Tchetgen, and Ted Cohen.
\newblock Negative controls: a tool for detecting confounding and bias in
  observational studies.
\newblock \emph{Epidemiology (Cambridge, Mass.)}, 21\penalty0 (3):\penalty0
  383, 2010.

\bibitem[Little and Rubin(2015)]{littlebook}
Roderick~JA Little and Donald~B Rubin.
\newblock \emph{Statistical analysis with missing data}.
\newblock John Wiley \& Sons, 2015.

\bibitem[Miao et~al.(2018)Miao, Geng, and
  Tchetgen~Tchetgen]{miao2018identifying}
Wang Miao, Zhi Geng, and Eric~J Tchetgen~Tchetgen.
\newblock Identifying causal effects with proxy variables of an unmeasured
  confounder.
\newblock \emph{Biometrika}, 105\penalty0 (4):\penalty0 987--993, 2018.

\bibitem[Ogburn and Vanderweele(2012)]{ogburn2012bias}
Elizabeth~L Ogburn and Tyler~J Vanderweele.
\newblock Bias attenuation results for nondifferentially mismeasured ordinal
  and coarsened confounders.
\newblock \emph{Biometrika}, 100\penalty0 (1):\penalty0 241--248, 2012.

\bibitem[Robins et~al.(2000)Robins, Rotnitzky, and
  Scharfstein]{robins2000sensitivity}
James~M Robins, Andrea Rotnitzky, and Daniel~O Scharfstein.
\newblock Sensitivity analysis for selection bias and unmeasured confounding in
  missing data and causal inference models.
\newblock In \emph{Statistical models in epidemiology, the environment, and
  clinical trials}, pages 1--94. Springer, 2000.

\bibitem[Rosenbaum and Rubin(1983)]{rosenbaum1983assessing}
Paul~R Rosenbaum and Donald~B Rubin.
\newblock Assessing sensitivity to an unobserved binary covariate in an
  observational study with binary outcome.
\newblock \emph{Journal of the Royal Statistical Society: Series B
  (Methodological)}, 45\penalty0 (2):\penalty0 212--218, 1983.

\bibitem[Shah and Peters(2018)]{shah2018hardness}
Rajen~D Shah and Jonas Peters.
\newblock The hardness of conditional independence testing and the generalised
  covariance measure.
\newblock \emph{arXiv preprint arXiv:1804.07203}, 2018.

\bibitem[Sharma et~al.(2018)Sharma, Hofman, Watts, et~al.]{sharma2018split}
Amit Sharma, Jake~M Hofman, Duncan~J Watts, et~al.
\newblock Split-door criterion: Identification of causal effects through
  auxiliary outcomes.
\newblock \emph{The Annals of Applied Statistics}, 12\penalty0 (4):\penalty0
  2699--2733, 2018.

\bibitem[Wang and Blei(2019)]{wang2019multiple}
Yixin Wang and David~M Blei.
\newblock Multiple causes: A causal graphical view.
\newblock \emph{arXiv preprint arXiv:1905.12793}, 2019.

\end{thebibliography}

\end{document}